\documentclass[conference]{IEEEtran}
\IEEEoverridecommandlockouts
\usepackage{cite}
\usepackage{amsmath,amssymb,amsfonts}
\usepackage{algorithmic}
\usepackage{graphicx}
\usepackage{textcomp}
\usepackage{xcolor}
\usepackage{hyperref}
\def\BibTeX{{\rm B\kern-.05em{\sc i\kern-.025em b}\kern-.08em
    T\kern-.1667em\lower.7ex\hbox{E}\kern-.125emX}}
\begin{document}

\title{Energy-based Accounting Model for Heterogeneous Supercomputers}

\author{\IEEEauthorblockN{Cristian Di Pietrantonio}
\IEEEauthorblockA{
\textit{Pawsey Supercomputing}\\ \textit{Research Centre}\\
Perth, Australia \\
cdipietrantonio@pawsey.org.au}
\and
\IEEEauthorblockN{Christopher Harris}
\IEEEauthorblockA{
\textit{Pawsey Supercomputing}\\ \textit{Research Centre}\\
Perth, Australia \\
chris.harris@pawsey.org.au}
\and
\IEEEauthorblockN{Maciej Cytowski}
\IEEEauthorblockA{
\textit{Pawsey Supercomputing}\\ \textit{Research Centre}\\
Perth, Australia \\
maciej.cytowski@pawsey.org.au}
}

\maketitle
\begin{abstract}
In this paper we present a new accounting model for heterogeneous supercomputers. An increasing number of supercomputing centres adopt heterogeneous architectures consisting of CPUs and hardware accelerators for their systems. Accounting models using the \emph{core hour} as unit of measure are redefined to provide an appropriate charging rate based on the computing performance of different processing elements, as well as their energy efficiency and purchase price. In this paper we provide an overview of existing models and define a new model that, while retaining the core hour as a fundamental concept, takes into account the interplay among resources such as CPUs and RAM, and that bases the GPU charging rate on energy consumption. We believe that this model, designed for Pawsey Supercomputing Research Centre's (Pawsey) next supercomputer Setonix, has a lot of advantages compared to other models, introducing carbon footprint as a primary driver in determining the allocation of computational workflow on heterogeneous resources.

\end{abstract}
\section{Background}
%

Computer-enabled investigations such as simulations and big data analysis count an ever-increasing number of calculations, requiring an outstanding amount of resources to execute \cite{CIELO2021102758}. Two items make up the cost for a researcher to run computations: the time required for the computation to run to completion and the compute bandwidth of the system employed. The two factors are often one inversely proportional to the other: the more compute bandwidth is available, the lesser time is required for the computation to complete. Because time is a very scarce resource researchers have no control over, supercomputing centres around the world hold a fundamental role in scientific discovery by providing meritorious researchers access to supercomputers. A supercomputer is a complex computing system comprising thousands of high performance compute nodes, each hosting multi-core CPUs, operational memory and possibly hardware accelerators, interconnected through a high-bandwidth low-latency network and provided with access to high performance file system. It represents a large pool of computational capacity many researchers can draw from to execute their workflows at scale within a reasonable time frame. Typically, a job scheduler is assigned the task to collect user requirements, schedule and execute user defined jobs on a supercomputer as well as gather usage information. User requirements are simply the description of resources needed to support the execution of the job, starting from number of compute nodes down to detailed description of number of cores, memory and accelerators required within each node. 
 
Even though much larger than a common workstation, the capacity of a supercomputer is finite. Once the compute bandwidth is saturated, the waiting time increases. For this reason, a supercomputing centre often quantify the amount of consumable supercomputing resources available every year and define a merit allocation process that dictates who gets access to a portion of it. The solution to the first problem starts with the following definition. The consumable supercomputing resource is the hourly usage of CPU cores and its unit of measure is the Service Unit (SU). Until recently, the main hardware component providing compute bandwidth has been the CPU, whose atomic computational unit is the CPU core. Even though this is not the case anymore, the hourly CPU core usage provides researchers a familiar point of reference to measure and estimate their computational requirements. The cost of running a job can be expressed in service units and it is equivalent to the number of CPU cores allocated to the job multiplied by the elapsed time. A supercomputing centre starts from the given definition to come up with the total amount of Service Units available every year on its supercomputer.

Researchers estimate the amount of service units required to execute their computationally intensive workflows and apply for accessing the supercomputer. They provide a detailed description of their projects for supercomputing specialists to score the scientific impact and the technical ability to take advantage of supercomputing resources. Finally, available service units are allocated to those projects that scored highest.

Researchers plan their simulations and workflows in a way that allows them to optimise their service units usage. In particular, very often they will decide to choose the number of cores used by the job based on software's parallel scalability. Hence, the choice of the accounting and service units model can be seen as a strategic decision for the supercomputing centre as it also determines how the system is being utilised.


\section{Motivation}
Pawsey Supercomputing Research Centre is going to replace the current supercomputing system, based on homogeneous CPU-only compute nodes, with Setonix, a heterogeneous HPE Cray EX system featuring both AMD Milan CPUs and AMD Instinct GPUs. The same transition is happening in other supercomputing facilities around the world, and brought to light Europe's latest supercomputers LUMI \cite{LUMI} and Dardel \cite{Dardel} as well as US's first exascale systems Frontier \cite{Frontier} and El Capitan \cite{ElCapitan}. Similarly to a design methodology developed at Lawrence Berkeley National Laboratory \cite{Austin}, only subset of Setonix's compute nodes feature Graphics Processing Units (GPUs), specialised hardware that provides a dramatic increase in computational throughput for specific operations. This subset is referred to as \emph{GPU partition}, while the rest of the nodes form the \emph{CPU partition}. It can be assumed, that the number of CPU cores per node does not vary much across partitions. However, because a GPU is very different from a CPU, the conveniently simple core hour model cannot be easily extended by identifying a core equivalent within the accelerator. On one hand, a GPU can only be reasonably used to solve computational problems with determined characteristics, whereas a CPU supports a much wider array of computer programs. The increasing computational bandwidth of modern GPUs is becoming harder to saturate, too. On the other hand, by limiting itself to particular use cases, a GPU can achieve a much higher computational throughput than a CPU, with greater energy efficiency.

To accommodate the increasing variety of computational workflows and improve resource utilisation, the compute nodes of the new supercomputer Setonix can host multiple jobs, a configuration often called \emph{shared access}. This is in contrast with \emph{exclusive node access}, a modality the compute nodes of Pawsey's current supercomputer Magnus are configured in, where each node can only support one job at a time. 

Because of the radical changes introduced with the new supercomputer Setonix, and the rising importance of energy efficiency from the point of view of monetary cost and impact on environment, we quickly realised that existing accounting models wouldn't have worked anymore and we started searching a new one. Among our discoveries is that an accounting model can model and lead to the optimisation of the energy efficiency of workflows running on a supercomputer. Researchers applying for supercomputing resources or planning their workflows usually consider service units usage (cost) versus time to solution (speed). In this paper we show that by linking service units with energy consumption, researchers' scheduling decisions result in a positive impact on the {carbon footprint} of their jobs because our proposed accounting model indirectly constrains them to use the most energy efficient computational resource, within a heterogeneous system, for the job at hand. This outcome aligns well with the positive and important push in the research community to further reduce carbon footprint of computations \cite{Stevens}. We strongly believe that using accounting models similar to the one presented in this paper can help supercomputing centres to achieve their performance goals and optimise energy efficiency at the same time.

Our objectives are to present our new accounting model for heterogeneous supercomputing systems based on energy consumption and, by closely describing the process of its definition, to provide guidance for centres looking to define their own. The paper is structured as follows. We begin with presentation of existing accounting models, including those developed for heterogeneous systems. We then introduce our model through a mathematical formulation and describe its main features. Next, we use application benchmarking data to compare various models and to discuss benefits of energy-based models. We finalise the paper by formulating conclusions.


\section{Existing Accounting Models}

The process of finding a proper accounting model for the new system starts with the exploration of the ones currently adopted, or used in past, at various supercomputing centres. The objective is to learn from them both valuable insights and shortcomings that can be avoided. 
An accounting model heavily depends on the characteristics of the supercomputer it is applied to. To narrow down the search to models taking into account the most common and relevant aspects, we have asked ourselves the following questions.

\begin{enumerate}
    \item {\bf Can multiple jobs share the same node?} A compute node can be configured in either one of modes listed below.
    \begin{itemize}
    \item {\it Exclusive access} - a job is allocated and charged for resources at compute node granularity. In other words, when a job is scheduled for execution on one or more nodes, it gets assigned all of their resources for the entire duration of the run, even when only a fraction of them are actually needed.
    \item {\it Shared access} - a job is allocated and charged for resources at a fine grained level on each requested node. In this configuration, a node that already hosts a job running on a subset of its resources will still allow the remaining ones to be assigned to other jobs. 
    \end{itemize}
    \item {\bf What type of system architecture is modelled?} Authors have identified three pertinent answers to this question.
    \begin{itemize}
        \item {\it Single CPU architecture system} - each compute node has the same model of CPU, or the model does not distinguish between different types of CPU core and compute node. They all amount to the same quantity of service units.
        \item {\it Multiple CPU architectures system} - the model distinguishes among different types of CPU core and compute node that may be present on the system by applying different charging factors;
        \item {\it Heterogeneous system} - the accounting model is aware of the presence of computing devices of different nature, most commonly CPUs and GPUs. This is what makes a system \emph{heterogeneous}. The presence and use of other {\it specialised hardware}, such as fast node-local storage, can be also taken into account for billing purposes.
    \end{itemize}
\end{enumerate}

Pawsey's next generation supercomputer Setonix is a heterogeneous system (with CPU and GPU nodes) whose nodes are configured for shared access. This allows the centre to implement a relatively flexible configuration of the system with the goal of supporting compute and memory requirements of various workflows. 

What follows is a description of few already established accounting models, with references. Differences between node exclusive and node sharing scenarios are discussed first. Various approaches to accounting on heterogeneous systems is discussed next. We clearly identify few typical approaches which are used later in Section \ref{sec:comparison} for comparison. 

Our intention is not to criticise or discourage specific approaches to accounting. We make an assumption that each supercomputing centre develops its own model based on important strategic decisions which are not known for the authors and as such are beyond discussion and scope of this paper. 

\subsection{Node Sharing}

In this subsection, we focus on comparing exclusive and shared node access accounting models. We provide few examples which are an important context for node sharing accounting proposed in our model.

\subsubsection{Exclusive Access}

There are numerous examples of supercomputing systems configured for granting jobs exclusive access to nodes, and their accounting models are formulated considering the whole node as an integral resource. One of such examples is the model developed for the Archer2 system \cite{Archer2}, which defines a Compute Unit (CU) as 1 hour use of a 128-core AMD EPYC node. It follows that, in cases like this one, researchers are asked to estimate their allocation needs in \emph{node hours} rather than core hours. 

A slightly different approach is implemented at Pawsey for Magnus, a Cray XC40 system. Magnus 
also mandates exclusive access to its nodes, making finer grained resource allocation not possible; however, the accounting model is based on the equivalence between 1 Service Unit and 1 CPU core hour. In short, one hour use of a single 24-core compute node of the current supercomputer Magnus is equivalent to 24SU. There is an advantage in billing for core hours, even on a system like this, when a supercomputing site may operate more than a single HPC system: the core hour is the minimum common denominator across all of them. The adoption of a single model keeps user experience uniform across systems.


\subsubsection{Shared Access}
\label{shared_nodes}
The difference between accounting models for exclusive and shared nodes is very well presented in the case of the Finnish IT Centre for Science (CSC) computational resources \cite{Puhti}. CSC's billing units (BU) are defined for each consumable resource of shared nodes of the Puhti system, whereas the BU equivalence of a single node usage is defined for exclusive nodes of the Mahti system. 

The total BU consumption for a compute job on Puhti (shared access) depends linearly on the number of requested consumable resources. BU rates for Puhti are defined in the following way:
\begin{itemize}
    \item Each reserved core consumes 1 BU per hour,
    \item Each GiB of reserved memory consumes 0,1 BU per hour, 
    \item Each reserved GiB of NVMe disk (if available) consumes 0,006 BU per hour,
    \item Each reserved GPU consumes 60 BU per hour.
\end{itemize}       
In fact, as we can see, Puhti is a heterogeneous system. Puhti's CPU to GPU billing units ratio will be discussed separately in the next subsection. Here, we only look at how node sharing is accounted for. The total BU consumption can be estimated by multiplying walltime of the job in hours, by the sum of BU associated with all consumed resources:
\begin{multline*}
(1\cdot \emph{NCores} + 0,1\cdot\emph{MemGiBs} + 0,006\cdot\emph{NVMeGiBs} \\ + 60\cdot\emph{NGPUs}) \cdot \emph{Walltime\_hours}.
\end{multline*}

Moreover, Puhti's model includes billing policy for increased quota for the \emph{scratch} filesystem and project storage. 
\subsection{System architecture}

In this subsection, we focus on heterogeneity of supercomputing systems and its implications to accounting. From authors' perspective, the most interesting scenario is when the supercomputing system is composed of both CPU and GPU architecture. The approaches presented here will be later used for comparison analysis in Section \ref{sec:comparison}.

\subsubsection{Multiple CPU Architectures Systems}
Supercomputers with multiple CPU architectures are built as a result of:
\begin{itemize}
\item targeted design, to address specific requirements of two or more types of computational workflows, or
\item upgrades which involve deployment of newer, more computationally and/or energy efficient, architectures.
\end{itemize} 
In both cases, supercomputing centres usually define different charge factors for partitions related to various CPU architectures. As an example, The National Energy Research Scientific Computing Center's (NERSC) Cori system is based on Intel Haswell and Intel KNL architectures \cite{cori}. Cori's charge factors favour codes that can benefit from many core design of the Intel KNL CPU and its AVX512 vector instructions, with per single Node-Hour charge of 80 for Cori-KNL and 140 for Cori-Haswell. This model is in general accordance with an approach that usage of more energy efficient cores should be rewarded.

On the contrary, linking accounting model directly with peak performance or peak performance ratio between CPU architectures, very often leads to undesirable consequences. Researchers running workflows that cannot reach the full performance advantage of the new architecture might decide to use less energy efficient solutions based on service units cost.

For instance, introducing a service units boost (or multiplier) related to CPUs ability to support longer vector lengths might mislead researchers to use less energy efficient solutions in their workflows. An example is introducing 2x multiplier between AVX512 and AVX2 CPU core architectures. Although selected scientific codes can leverage AVX512 instructions, especially when using high performance linear algebra libraries, the performance benefits are usually much smaller than 2x speedup. In addition, as it was shown in \cite{Guermouche}, the energy efficiency (measured in watts per core hour) does not necessarily increase with the use of longer vector instructions available in AVX512. It follows that researchers should not be penalised for using AVX512 cores even if their computational workflows' speedup per core is minimal when compared to AVX2 architectures.


\subsubsection{Heterogeneous Systems}

Heterogeneous supercomputers are systems which are composed of CPU and accelerator or special purpose architectures (e.g. machine learning processors). Historically, one of the first CPU/GPU accounting models was introduced at Oak Ridge Leadership Computing Facility (OLCF). Decommissioned in 2019, OLCF's Titan \cite{Titan} was a heterogeneous Cray XK7 system with peak performance exceeding 27 PetaFLOPS. Titan made its Top500 \cite{Top500} debut in 2012 at 1st spot and remained on Top 10 for twelve consecutive editions of the list. Most importantly, in the context of this paper, Titan was the first hybrid, GPU accelerated systems of that scale. It combined 16-core AMD Opteron processors with NVIDIA Kepler GPUs. \\
Titan's accounting model was based on equivalence between 1 Service Unit and 1 CPU core hour, additionally introducing an equivalence between a 1 Service Unit and 1 GPU streaming multiprocessor (SM). As a result, 1 hour usage of a single Titan's node was worth of 30 Service Units (16 AMD Opteron cores plus 14 NVIDIA Kepler's SM). It should be noted, that node sharing was not allowed on Titan and users were charged for both GPU and CPU usage even in the case of non-GPU-ready application. Titan's accounting model can be considered as revolutionary, however it cannot be easily transferred to current GPU architectures composed of 100+ SMs. Titan model will be one of the reference models in the comparison presented in Section \ref{sec:comparison}.

Accelerators are also a very good example to illustrate why accounting models should not be directly linked with peak performance. Modern GPU accelerators' double precision floating point (DP FLOPs) performance is 200x higher compared to DP FLOPS performance of single cores of modern CPU architectures\footnote{Comparing NVIDIA A100 architecture (9.7 DP TFLOPs) with AMD EPYC 64-core Milan 7763 (2.5 DP TFLOPs).}. If we were to base the GPU charging rate on the ratio between GPU peak performance and the one of a single CPU core, then we would be using a factor that high. Achieving performance close to theoretical peak on accelerators requires highly optimised algorithms and appropriate programming techniques. As a result, most of the simulation packages struggle to reach even 75\% of peak DP FLOPS performance at scale \footnote{OLCF's Summit system achieves 74\% of peak performance on High Performance Linpack benchmark \cite{Top500}}. It follows that accounting models based on peak performance can discourage energy efficient GPU workflows, this is discussed in more details in Section \ref{sec:comparison}.

The CSC's Puhti node sharing accounting model discussed in previous subsection is an interesting heterogeneous model which provides a good introduction to our model. Puhti is based on Intel Xeon Gold 6230 ``Cascade Lake'' and NVIDIA V100 with NVLink architectures. Each reserved GPU consumes 60 BU per hour, being an equivalent of 60 CPU core hours. We note that:
\begin{itemize}
    \item Intel Xeon Gold 6230 ``Cascade Lake'' (20-core) Thermal Design Power (TDP) is 125W, i.e. 6.25W per core, and 
    \item NVIDIA V100 with NVLink TDP is 300W.
\end{itemize}
Each of Puhti's GPU nodes is composed of 40 Intel Xeon Gold 6230 ``Cascade Lake'' cores and four NVIDIA V100 GPUs. Single hour use of one GPU device together with $1/4$ of CPU cores per node is equivalent, from TDP perspective, to $$300Wh + \frac{1}{4} \cdot (2 \cdot 125Wh) = 362.5Wh.$$ This is exactly $58x$ more than per CPU core $6.25Wh$ figure and might be an explanation of CSC's approach. We note that $60BU$ charge factor introduced by CSC is very similar to the energy-based accounting approach proposed in this paper.      

\section{A model based on energy efficiency}
%
%

The previous sections have provided an overview of past and present charging schemes adopted at various supercomputing centres accompanied by a discussion of their points of strength and weaknesses. Drawing from the author's lessons learnt and experience in running a supercomputing facility, we now introduce a new accounting model. It aims at covering important gaps in current models, and promoting energy efficiency as driving motivation for deciding what type of hardware is best suited to run a given job. A second motivation is that the model must be simple enough for users to be able to predict what the cost of a job will be, and to estimate the size of their computational needs when applying for allocation rounds.

\subsection{Model overview}
Mathematically, an accounting model is a function of several variables that associates a cost to a given job as a number of service units. The variables capture important characteristics of a job such as the number of compute nodes it spanned and, on each one of them, the number of CPU cores, GPUs and amount of memory utilised. More formally, let
\begin{itemize}
\item $n$ be the number of compute nodes requested,
\item $w$ a weight based on the type of nodes used,
\item $\mathbf{f}_c$ the fraction of CPU cores requested on every node,
\item $\mathbf{f}_g$ the fraction of GPUs requested on every node,
\item $\mathbf{f}_m$ the fraction of memory requested on every node, and
\item $t$ the time it took for the job to complete, expressed in hours.
\end{itemize}

Then, the cost function \emph{cost}, representing the new accounting model, is defined as
\begin{equation}\label{eq:accounting_model}
    \emph{cost}(n, w, t, \mathbf{f}_c, \mathbf{f}_g, \mathbf{f}_m) = wt \sum_{i=1}^n\emph{max}(f^i_c, f^i_g, f^i_m).
\end{equation}

For every node, the largest fraction of resource used represents how much of that node has being allocated to the job. A compute node of the new supercomputer Setonix can host multiple jobs at the same time, provided that there are enough resources available; hence, a job is only charged for the portion of computing resources of a node it requested. In contrast, other supercomputers adopts the \emph{exclusive node access} configuration that allocates (and charges for) entire nodes irrespective of whether all resources within them are actually consumed by a job.

A job may be heterogeneous. That is, it may use different resources, possibly in different amounts, on the compute nodes allocated to it. The cost function accounts for it by taking the maximum fractional usage over all resources independently for every compute node. The maximal fractional usage for each nodes are then added to together. The result is finally scaled by the partition weighting factor, explained in later subsections, and multiplied by the job execution time.

\subsection{One core hour, one service unit}
Compute nodes of a supercomputer are usually partitioned according to whether they provide GPU acceleration. For the \emph{CPU partition}, the weight $w$ is equal to $C$, the number of CPU cores available on a compute node in the partition (without loss of generality, and as it is almost always the case, we can assume compute nodes in the same partition are identical). Because there are no GPUs, $\mathbf{f}_g = \mathbf{0}$. In the case when $\mathbf{f}_m \leq \mathbf{f}_c$, that is, the job is compute intensive rather than memory intensive, the cost reduces to

\begin{align}
    \emph{cost}(n, w, t, \mathbf{f}_c, \mathbf{f}_g, \mathbf{f}_m) &= tC\sum_{i=1}^nf^i_c\\
    &= tn_c
\end{align}

where

\begin{align}
    n_c=C\sum_{i=1}^nf^i_c.
\end{align}

Notice the particular case when $n = C = t = 1$. It is the familiar equivalence between 1 Service Unit and 1 hour use of a CPU core. Traditional CPU-only, computational intensive jobs are charged as many service unit as the core hours they spent running on the supercomputer, retaining the simple to use yet most adequate, commonly agreed on accounting model for this type of work.

\subsection{Taking into account memory}
The contribution by node $i$ to the total cost when $f^i_m > f^i_c$ and $f^i_g = 0$ depends on the amount of memory used. The reason is that when a job consumes proportionally more memory than CPU cores on a node, it hinders the ability of other jobs to use the remaining cores. In the extreme, one core using all the memory on the node would prevent other jobs from running on the node at the same time. The cost of such jobs is

\begin{align}
    \emph{cost}(n, w, t, \mathbf{f}_c, \mathbf{f}_g, \mathbf{f}_m) &= tC\left(\sum_{j : j\neq i}\emph{max}(f^j_c, f^j_g, f^j_m) + f^i_m\right)\\
    &= tC\sum_{j : j\neq i}\emph{max}(f^j_c, f^j_g, f^j_m) + te_c.
\end{align}

where

\begin{equation}\label{eq:mem_to_cores}
e_c=C f^i_m
\end{equation}
maps the amount of memory used on node $i$ to an equivalent number of cores for charging purposes. The value of $e_c$ might not be integral because, in general, $f^i_m \in (0, 1]$. For ease of interpretation, we would like $e_c$ to take on integer values in $\{1, \ldots, C\}$. Let $M_{\emph{tot}}$ be the total amount of memory available on a node, and $M^i_{\emph{used}}$ be the quantity of memory used by a job on node $i$. We can think of distributing the available memory equally to each  core. Let $M_{\emph{core}}$,
\begin{equation}
    M_{\emph{core}} = \frac{M_{\emph{tot}}}{C},
\end{equation}

be the amount of memory associated to each core. Ideally, this is the maximum amount of memory a core can consume without unfairly impacting tasks running on other CPU cores. Then, we can define $f^i_m$ as
\begin{equation}\label{eq:fm_discretised}
    f_m^i = \left\lceil\frac{M^i_{\emph{used}}}{M_{\emph{core}}}\right\rceil.
\end{equation}

Indeed, $f^i_m$ now is a step function having a finite set of fractional numbers as co-domain. With Equation~\ref{eq:fm_discretised} in place, $e_c$ takes on values in the same set of possible values for the number of CPU cores that might be used on a compute node.

\subsection{GPU accounting based on energy consumption}


The point of strength of the accounting model presented in this paper is its ability of reconciling the traditional definition of Service Unit with the increasing priority of optimizing energy efficiency when charging for GPU usage. A GPU provides more computational power than a CPU and consumes more energy, so it makes sense to charge more for its use. A GPU is also more energy efficient, meaning it provides more computational bandwidth per unit of energy, so its use must be encouraged when a program can take significant advantage of it. The definition of the weight for the GPU partition must take into account these factors.

The first step is to find a conversion rate $R$ between Watt and Service Unit. The hourly use of an entire CPU with $c$ cores costs $c$ Service Units. Since the TDP $E_c$ of the CPU is known, we can derive the cost $R$ in Service Units of using 1 Watt of power per hour:
\begin{equation}
    R = \frac{c}{E_c}.
\end{equation}

Given the TDP $E_g$ of a GPU, the cost $G$ of its hourly power consumption in terms of Service units is
\begin{align}
G &= R E_g\\
&= \frac{E_g}{E_c}c\label{eq:energy_consumption_in_su}\\
&= \alpha c
\end{align}

where

\begin{align}
    \alpha = \frac{E_g}{E_c}.
\end{align}

In our model, the cost of one hour use of a GPU is equal to the hourly Service Unit consumption of the CPU of reference scaled by a factor representing how much more energy the GPU consumes compared to that CPU.

The reasoning can be extended to an entire compute node belonging to the GPU partition. Let $\sum_{i}E^i_c$ now be the cumulative TDP over all CPUs and $C = \sum_i c_i$ the total number of cores on that node, and $\sum_jE^j_g$ the total TDP of its GPUs; then, the hourly service unit consumption due to all GPUs is
\begin{equation}\label{eq:weight_gpu_part}
    w = w_g = \frac{\sum_jE^j_g}{\sum_{i}E^i_c}C.
\end{equation}
GPUs consume more service units than CPUs because $E_g > E_c$, and anyone using compute nodes with GPUs should be using GPU accelerated codes, and will be charged as if they were. For this reason, equation \ref{eq:weight_gpu_part} defines also the weight $w$ assigned to the GPU partition. The fraction $f^i_g$ of GPU used on node $i$ is simply the number of GPUs requested over the number of GPUs available.

The cost of using a GPU does not depend on its peak computational throughput, which approximately an order of magnitude larger than the CPU. The reason is twofold. From the user perspective, many scientific codes would overpay, meaning they would be charged much more in proportion to what they can get, not being able to fill up the capacity of modern accelerators. 

From the author's perspective, we would like to avoid users using CPUs to minimise service unit billing, at the cost of increasing the execution time and energy consumption, leading to the GPU partition being underutilised and a reduced job throughput for the overall supercomputer. However, many popular supercomputing applications experience a significant speedup when using accelerators. Therefore, we strive to find an appropriate weight such that there is an incentive for users in using GPUs while, at the same time, energy consumption is appropriately accounted for. We show in the following Section~\ref{sec:comparison} that equation \ref{eq:weight_gpu_part} is an appropriate choice.

Note that in our model we intentionally decide to use TDP rather than real (measured) energy consumption of a job. This allows us to construct a predictive model in which SU yearly budget can be estimated. This approach is also similar to charging for the number of cores requested by the job instead of the number of actually used cores. It is up to who submits the job to make sure that the requested resource is fully utilised. We also assume that CPU and GPU power consumption is the main contributor to the total power consumption of given computational job.   


\subsection{Extending the model}
A GPU is not the only new type resource being integrated into a modern supercomputer. For instance, \emph{near node} storage gives I/O intensive applications access a fast, although limited, disk space where to write temporary files for the entire duration of a job. Its use, much like the one of RAM, should be charged for as it is a consumable resource. One can easily extend our model to account for the use of other resources on a per node basis by adding more arguments to the \emph{max} function in equation \ref{eq:accounting_model}.

\section{Comparison and Discussion}
\label{sec:comparison}


To explore the energy-based accounting model described in the previous section, we apply it to a hypothetical test system based on performance results presented on the NVIDIA HPC Application Performance web page~\cite{NVIDIAHPC}. 
In those results, the CPU benchmarks were conducted on nodes with dual Xeon Gold 6240 18-core processors, and the GPU benchmarks used four A100 SMX GPUs each with 108 streaming multiprocessors (SMs).
The theoretical 64 bit precision floating point operations per second (FLOPs) fused multiply-add (FMA) performance for the CPU and GPUs are provided in Table~\ref{tab:psumm} below, along with the TDP. The figure for the CPU was calculated from the base clock frequency of 2.6 GHz and 18 cores each with two AVX-512 FMA units~\cite{XeonTDP}, and the GPU value is the FMA performance published by NVIDIA~\cite{A100TDP}. 

\begin{table}[h]
    \centering
    \begin{tabular}{|l|c|c|c|c|}\hline
    \textbf{Processor} & Cores & SMs & FLOPs & TDP\\\hline 
    Xeon Gold 6240     & 18 & - & 1.5e12 & 150W \\
    A100 SMX           & - & 108 & 9.7e12 &400W \\\hline
    \end{tabular}
    \caption{Processor Summary}
    \label{tab:psumm}
\end{table}

Using this test system, we now explore the application of the SM-based, performance-based, and energy-based accounting models. For each of these, the service unit is normalised to one core-hour of the CPU node. We then calculate the number of cores on the CPU node, $C$ as follows:

\begin{equation}
C = 2 \cdot 18 = 36.
\end{equation}

This results in a CPU node hour costing 36 SU. Using this cost for a CPU node, the various costs for a GPU node for each example model are explored in the following subsections.

\subsection{Example SM-Based Model}

In an SM-based accounting model the charge a for GPU node hour, $w_g$, is the number of streaming multiprocessors in the GPU nodes. For the example system above, this is as follows:

\begin{equation}
    w_g = 4 \cdot 108 = 432\\
\end{equation}

Based on the node equivalence provided for the benchmarks, the cost of running each benchmark for one hour, and the cost ratio between the two types of nodes can be calculated as shown in Table~\ref{tab:sm_costs}.

In the tables presented in this section, there is a row for each application benchmark. The number of CPU nodes needed to provide equivalent performance to one GPU node that was published by NVIDIA is provided in the {\emph{Perf. Ratio}} column. The SU cost on running the application for one hour on the equivalent CPU nodes is provided in the {\emph{CPU Charge}} column. The SU cost of running the application for one hour on the GPU node is provided in the {\emph{GPU Charge}} column. The ratio between these two is obtained by dividing the CPU charge by the GPU charge and provided in the {\emph{Cost Ratio}} column.

\begin{table}[h]
    \centering
    \begin{tabular}{|l|l|c|c|c|c|}\hline
                     & Perf. & CPU    & GPU    & Cost \\
         Application & Ratio & Charge & Charge & Ratio\\\hline
         FUN3D            & 41  & 1,476 & 432 & 3.42 \\
         RTM              & 32  & 1,152 & 432 & 2.67 \\
         SPECFEM3D        & 105 & 3,780 & 432 & 8.75 \\
         AMBER            & 153 & 5,508 & 432 & 12.7 \\
         GROMACS          & 23  & 828   & 432 & 1.92 \\
         LAMMPS           & 59  & 2,124 & 432 & 4.92 \\
         NAMD             & 36  & 1,296 & 432 & 3.00 \\
         Relion           & 12  & 432   & 432 & 1.00 \\
         GTC              & 53  & 1,908 & 432 & 4.42 \\
         MILC             & 108 & 3,888 & 432 & 9.00 \\
         Chroma           & 99  & 3,564 & 432 & 8.25 \\
         Quantum Expresso & 13  & 468   & 432 & 1.08 \\
         ICON             & 15  & 540   & 432 & 1.25 \\\hline
    \end{tabular}
    \caption{SM-based accounting for one hour benchmarks}
    \label{tab:sm_costs}
\end{table}

\subsection{Example Peak Performance-based Model}

For the investigation of a performance-based accounting model, we consider the theoretical 64 bit precision floating point operations per second (FLOPs) fused multiply-add (FMA) performance for the processors. Based on the processor performance detailed previously in Table~\ref{tab:psumm}, the theoretical performance for the CPU node, $T_c$, and GPU node, $T_g$, is:

\begin{align}
T_c &= 2 \cdot 1.5e12\\
    &= 3e12\\
T_g &= 4 \cdot 9.7e12\\
    &= 38.8e12
\end{align}

Using these values, we construct the example performance-based accounting model. This first uses the same 36 SU charge for the CPU node, and then uses the ratio of the performance for the GPU node as follows.

\begin{align}
     w_g &= \frac{T_g}{T_c} C\\
       &= \frac{(4 \cdot 9.7e12)}{(2 \cdot 1.5e12)} 36\\
       & \approx 466
\end{align}

This results in a charge of 466 per GPU node. Using this alternate accounting model we can generate the costs in a similar manner to before.

\begin{table}[h]
    \centering
    \begin{tabular}{|l|l|c|c|c|c|}\hline
                     & Perf. & CPU    & GPU    & Cost \\
         Application & Ratio & Charge & Charge & Ratio\\\hline
         FUN3D            & 41  & 1,476 & 466 & 3.17 \\
         RTM              & 32  & 1,152 & 466 & 2.47 \\
         SPECFEM3D        & 105 & 3,780 & 466 & 8.11 \\
         AMBER            & 153 & 5,508 & 466 & 11.8 \\
         GROMACS          & 23  & 828   & 466 & 1.78 \\
         LAMMPS           & 59  & 2,124 & 466 & 4.56 \\
         NAMD             & 36  & 1,296 & 466 & 2.78 \\
         Relion           & 12  & 432   & 466 & 0.93 \\
         GTC              & 53  & 1,908 & 466 & 4.09 \\
         MILC             & 108 & 3,888 & 466 & 8.34 \\
         Chroma           & 99  & 3,564 & 466 & 7.64 \\
         Quantum Expresso & 13  & 468   & 466 & 1.00 \\
         ICON             & 15  & 540   & 466 & 1.16 \\\hline
    \end{tabular}
    \caption{Performance-based accounting for one hour benchmarks}
    \label{tab:perf_costs}
\end{table}

\subsection{Example Energy-based Model}

For the energy-based accounting model we consider the ratio of TDP of the processors. Using the CPU and GPU TDP values from Table~\ref{tab:psumm} above, we use Equation~\ref{eq:weight_gpu_part} as follows

\begin{align}
     w &= \frac{\sum_{j}E^j_g}{\sum_{i}E^i_c}C\\
       &= \frac{(4 \cdot 400)}{(2 \cdot 150)} 36\\
       &\approx 192
\end{align}

This results in a GPU node hour costing 192 SU.

Based on the node equivalence provided for the benchmarks, the cost of running each benchmark for one hour and the cost ratio can be calculated as shown in Table~\ref{tab:costs}.

\begin{table}[h]
    \centering
    \begin{tabular}{|l|l|c|c|c|c|}\hline
                     & Perf. & CPU    & GPU    & Cost \\
         Application & Ratio & Charge & Charge & Ratio\\\hline
         FUN3D            & 41  & 1,476 & 192 & 7.69 \\
         RTM              & 32  & 1,152 & 192 & 6 \\
         SPECFEM3D        & 105 & 3,780 & 192 & 19.69\\
         AMBER            & 153 & 5,508 & 192 & 28.68 \\
         GROMACS          & 23  & 828   & 192 & 4.31 \\
         LAMMPS           & 59  & 2,124 & 192 & 11.06 \\
         NAMD             & 36  & 1,296 & 192 & 6.75 \\
         Relion           & 12  & 432   & 192 & 2.25 \\
         GTC              & 53  & 1,908 & 192 & 9.94 \\
         MILC             & 108 & 3,888 & 192 & 20.25 \\
         Chroma           & 99  & 3,564 & 192 & 18.56 \\
         Quantum Expresso & 13  & 468   & 192 & 2.44 \\
         ICON             & 15  & 540   & 192 & 2.81 \\\hline
    \end{tabular}
    \caption{Energy-based accounting for one hour benchmarks}
    \label{tab:costs}
\end{table}

\subsection{Discussion of the Example Models}

The energy-based accounting model provides an approximation for the relative power consumption of the benchmarks running on the two types of nodes. Based on the ratio in the final column of Table \ref{tab:costs}, all of the applications are likely to be more power efficient running using nodes with GPUs for this test system. 

However, the relative costs in the the final column of Table~\ref{tab:sm_costs} and~\ref{tab:perf_costs} have several codes that have similar or cheaper costs for the CPU nodes using those models. This may encourage researchers running similar workloads to make less energy efficient choices when scheduling work to make best use of their SU allocation.

There is also selection bias in the range of benchmarks presented here, in that they represent a subset of applications that run particularly well on GPU architectures. In real world supercomputing workloads, it is likely that a larger proportion of application workflows may fall in this situation for the SM-based and performance-based accounting models where it is more cost-effective from a SU perspective but less power efficient to run on CPU nodes. 

This can be actually well presented for a hypothetical application X. We assume that X takes 1 hour to run on Intel Xeon Gold 6240 node (36 cores). The GPU versions of X takes $t$ hours to run on GPU node with four NVIDIA A100 SMX GPUs.  The speedup for application X is given by:

\begin{equation}
s=\frac{1}{t}    
\end{equation}

We also assume that researcher makes decision about the execution of X based on SU comparison for CPU and GPU nodes for application X and chooses node type with lower SU charge. The maximal energy consumption of the job can be estimated as:
\begin{equation}
EC_{total} = 
\begin{cases}
    1 \cdot \sum_{i}E^i_c \text{ (Wh)},& \text{if } SU_{cpu} \leq SU_{gpu}\\
    \frac{1}{s} \cdot \sum_{j}E^j_g \text{ (Wh)}, & \text{otherwise}
\end{cases}
\end{equation}
or, using the TDP numbers for these specific architectures detailed in Table~\ref{tab:psumm}:
\begin{equation}
EC_{total} = 
\begin{cases}
    1 \cdot 300 \text{ (Wh)},& \text{if } SU_{cpu} \leq SU_{gpu}\\
    \frac{1}{s} \cdot 1600 \text{ (Wh)}, & \text{otherwise}
\end{cases}
\end{equation}

\begin{figure}[h]
\centering
\includegraphics[scale=0.3]{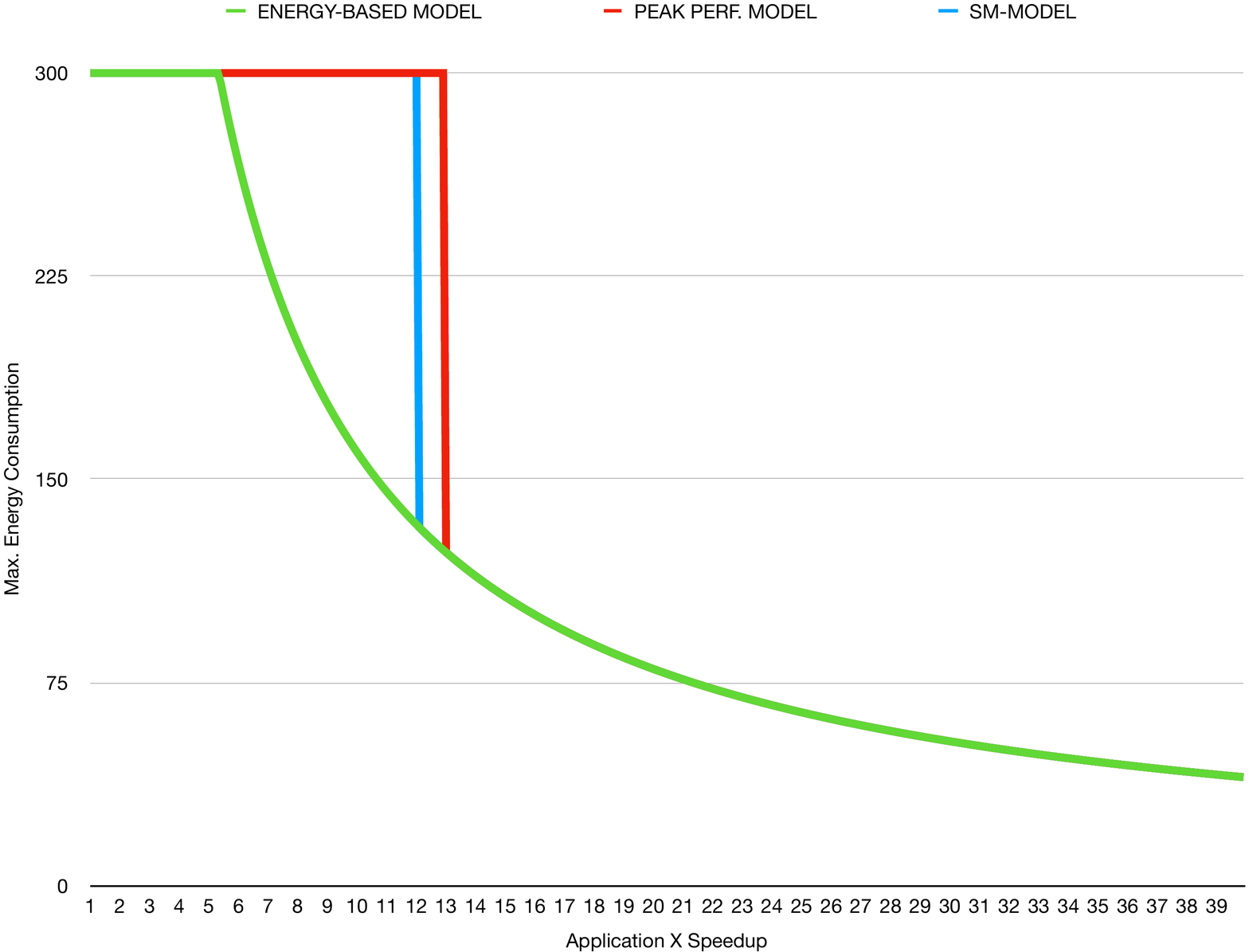}
\caption{Estimated maximal energy consumption of an application based on user decision plotted for 3 GPU service units models discussed in this paper.}
\label{fig:speedup}
\end{figure}

We can now compare $EC_{total}$ between three GPU service units models discussed in this section by additionally varying the speedup of the hypothetical application X. This is presented in Figure~\ref{fig:speedup}. We can see that for applications with speedup between 6 and 13 the energy-based model will result in the choice of more energy efficient node or partition.

\section{Conclusion}

The proposed model slightly changes the meaning of supercomputing accounting and service units. Although, in many cases it seems to be natural to link service units with peak performance, e.g. to justify the level of investment to funding institutions, we do not recommend it in practice. In the presented model, researchers optimising their service units usage will also optimise the cost and energy consumption (carbon footprint) of their computational jobs, workflows and projects. Therefore, researchers will be encouraged to choose the most energy efficient solution for their science. This approach discourages codes and workflows not optimised to take advantage of energy efficient accelerator architectures from running on GPU nodes, but it allows researchers and supercomputing support teams to identify and address accelerator optimisation opportunities and needs. Last but not least, researchers, supercomputing centres and funding institutions will benefit from the fact that service units in the presented model are directly linked with power consumption i.e. the operational cost of the supercomputing centre. 



\bibliography{biblio}

\begin{thebibliography}{10}

\bibitem{CIELO2021102758}
S.~Cielo, L.~Iapichino, J.~Günther, C.~Federrath, E.~Mayer, and M.~Wiedemann,
  ``Visualizing the world’s largest turbulence simulation,'' {\em Parallel
  Computing}, vol.~102, p.~102758, 2021.

\bibitem{LUMI}
EuroHPC, ``{LUMI} supercomputer.'' \url{https://www.lumi-supercomputer.eu},
  Accessed: 2021-08-16.

\bibitem{Dardel}
SNIC, ``Dardel supercomputer.''
  \url{https://www.pdc.kth.se/hpc-services/computing-systems/dardel-1.1053338},
  Accessed: 2021-09-01.

\bibitem{Frontier}
ORNL, ``Frontier supercomputer.'' \url{https://www.olcf.ornl.gov/frontier/},
  Accessed: 2021-09-01.

\bibitem{ElCapitan}
LLNL, ``El capitan supercomputer.''
  \url{https://www.hpe.com/au/en/compute/hpc/cray/doe-el-capitan-press-release.html},
  Accessed: 2021-09-01.

\bibitem{Austin}
B.~Austin, C.~Daley, D.~Doerfler, J.~Deslippe, B.~Cook, B.~Friesen, T.~Kurth,
  C.~Yang, and N.~J. Wright, ``A metric for evaluating supercomputer
  performance in the era of extreme heterogeneity,'' in {\em 2018 IEEE/ACM
  Performance Modeling, Benchmarking and Simulation of High Performance
  Computer Systems (PMBS)}, pp.~63--71, 2018.

\bibitem{Stevens}
A.~R.~H. Stevens, S.~Bellstedt, P.~J. Elahi, and M.~T. Murphy, ``The imperative
  to reduce carbon emissions in astronomy,'' {\em Nat Astron}, vol.~4,
  pp.~843–--851, Sept. 2020.

\bibitem{Archer2}
EPCC, ``Archer2 compute unit calculator.''
  \url{https://www.archer2.ac.uk/support-access/cu-calc.html}, Accessed:
  2021-08-25.

\bibitem{Puhti}
CSC, ``Puhti and mahti billing model.''
  \url{https://docs.csc.fi/accounts/billing/##puhti}, Accessed: 2021-08-23.

\bibitem{cori}
NERSC, ``Cori supercomputer.''
  \url{https://docs.nersc.gov/jobs/policy/#charge-factors-for-ay2021},
  Accessed: 2021-09-01.

\bibitem{Guermouche}
A.~Guermouche and A.-C. Orgerie, ``Thermal design power and vectorized
  instructions behavior,'' {\em Concurrency and Computation: Practice and
  Experience}, vol.~n/a, no.~n/a, p.~e6261.

\bibitem{Titan}
ORNL, ``What is a core hour on titan?.''
  \url{https://www.olcf.ornl.gov/2012/12/16/what-is-a-core-hour-on-titan/},
  Accessed: 2021-08-20.

\bibitem{Top500}
Top500.org, ``Top500 list,'' Accessed: 2021-08-20.

\bibitem{NVIDIAHPC}
NVIDIA, ``{HPC} application performance.'' \url{}, Accessed: 2021-08-24.

\bibitem{XeonTDP}
Intel, ``{Intel Xeon Gold 6240 Processor}.''
  \url{https://ark.intel.com/content/www/us/en/ark/products/192443/intel-xeon-gold-6240-processor-24-75m-cache-2-60-ghz.html},
  Accessed: 2021-08-24.

\bibitem{A100TDP}
NVIDIA, ``A100 gpu.'' \url{https://www.nvidia.com/en-au/data-center/a100/},
  Accessed: 2021-08-24.

\end{thebibliography}
\bibliographystyle{ieeetr}

\end{document}